\documentclass[journal=jctcce]{achemso}
\usepackage[utf8]{inputenc}
\usepackage{amsmath, bm}

\title{Symmetrized Drude Oscillator Force Fields Improve Numerical Performance of Polarizable Molecular Dynamics}
\author{Amro Dodin}
\email{adodin@lbl.gov}
\author{Phillip L. Geissler}
\affiliation{Chemical Sciences Division, Lawrence Berkeley National Laboratory, Berkeley, CA 94720,
United States.}
\alsoaffiliation{Department of Chemistry, University of California, Berkeley, CA 94720, United States.}
\date{\today}

\begin{document}

\begin{abstract}
    Drude oscillator potentials are a popular and computationally efficient class of polarizable models that represent each polarizable atom as a positively charged Drude core harmonically bound to a negatively charged Drude shell.
    We show that existing force fields that place all non-Coulomb forces on the Drude core and none on the shell inadvertently couple the dipole to non-Coulombic forces. 
    This introduces errors where interactions with neutral particles can erroneously induce atomic polarization, leading to spurious polarizations in the absence of an electric field and exacerbating violations of equipartition in the employed Carr-Parinello scheme. 
    A suitable symmetrization of the interaction potential that correctly splits the force between the Drude core and shell can correct this shortcoming, improving the stability and numerical performance of Drude oscillator based simulations.
    The symmetrization procedure is straightforward and only requires the rescaling of a few force field parameters.
\end{abstract}

\maketitle

\section{Introduction}
\label{sec:intro}

The success of classical molecular dynamics simulations hinges on our ability to construct simple, accurate models for the interactions between atoms.
Drude oscillator models are a widely used family of models that efficiently describe polarizable chemical species in Coulombically heterogeneous environments \cite{lemkul_empirical_2016}, such as a polarizable ion moving through a nanopore \cite{prajapati_computational_2020}, or from the bulk of an electrochemical cell into the double layer.
Each atom is split into a positively charged core particle and a negatively charged shell that are attached by a harmonic spring.
Since each atom is split into two particles, the non-Coulombic forces must also be split between the two particles.
Conventionally the entirety of the non-Coulombic force is assigned to the core and the shell particle evolves only under Coulombic forces.
In this paper, we show that this convention erroneously couples the atomic dipoles to non-electrostatic forces, leading to artificial atomic polarization induced by collisions with neutral particles.
This unphysical coupling exacerbates numerical issues that arise in Drude oscillator simulations including violations of equipartition and instabilities due to catastrophic polarization.
Fortunately, these errors can be easily corrected by symmetrizing how non-Coulombic forces are split between the Drude core and shell with minimal computational cost.

Inter-atomic interactions in classical molecular dynamics are described using empirical force fields that map a particular atomic configuration described by atomic positions, $\lbrace\bm{r_i}\rbrace$, and atomic charge distributions, $\lbrace\bm{x_i}\rbrace$, to its potential energy.
Atomic charge distributions may be described minimally, as a set of point charges $\lbrace\bm{x_i}\rbrace=\lbrace q_i\rbrace$, or may include higher order multipole moments to better capture electrostatic interactions.
The potential energy function is typically split into separate components,
\begin{equation}
    \label{eq:ff_struct}
    U(\lbrace \bm{r_i}\rbrace) = U_\mathrm{short}(\lbrace\bm{r_i}\rbrace) + U_C(\lbrace\bm{r_i}\rbrace, \lbrace \bm{x_i}\rbrace) + U_\mathrm{bond}(\lbrace \bm{r_i}\rbrace),
\end{equation}
where $U_\mathrm{short}$ is the short range (e.g. Lennard-Jones) interaction between atoms, $U_C$ describes their Coulomb interactions, and $U_\mathrm{bond}$ contains the interactions between bonded atoms (e.g. bond, angle, and dihedral potentials).
The functional form of each of these components and their parameters differ from one force field to the next and are fit to thermodynamic properties or electronic structure calculations of the systems of interest.

Most widely used force fields, including the OPLS \cite{jorgensen_development_1996}, AMBER \cite{wang_development_2004}, and CHARMM \cite{vanommeslaeghe_charmm_2010} general force fields as well as the TIPnP \cite{price_modified_2004, abascal_general_2005, rick_reoptimization_2004} and SPC \cite{berendsen_missing_1987} water models, employ a fixed point charge model for Coulomb interactions, which assigns a static point charge, $q_i$, to each atom.
The electrostatic component of the potential is given by
\begin{equation}
    \label{eq:UCfp}
    U_C^{(\mathrm{fpc})}(\lbrace \bm{r_i}\rbrace, \lbrace q_i\rbrace) = \frac{1}{2}\sum_{i\neq j}\frac{q_i q_j}{4\pi\epsilon_0 r_{ij}},
\end{equation}
where $\epsilon_0$ is the permittivity of free space and $r_{ij}$ is the distance between atoms $i$ and $j$.
The point charges are fit as parameters of the force field or assigned physically reasonable values (e.g. integer charges for simple atomic ions).
Fixed charge models are ubiquitous due to their ability to reasonably reproduce thermodynamic properties of homogeneous systems at low computational cost.
In particular, this type of model is referred to as  ``additive'' since the potential on each atom can be separately computed and summed together to obtain the total potential, allowing for straightforward computational acceleration through parallelization.

However, fixed charge models can fail to capture the properties of heterogeneous systems since they implicitly rely on a mean-field approximation in assuming that the atomic charge distributions do not depend on their local environments.
In reality, the charge distributions of atoms and molecules will change when they are placed in an electric field which in turn influences the Coulomb forces they exert on other atoms in the system.
This mean field assumption works well when most atoms experience similar electric fields, as is the case in homogeneous systems, but can yield inaccurate predictions when average  fields are spatially heterogeneous.
For example, fixed charge models can fail to capture ion-protein interactions at Ca$^{2+}$ binding sites \cite{li_representation_2015}, or the motion of ions between regions of different polarizability such as lipid-water interfaces \cite{yesylevskyy_polarizable_2010}.

Polarizable force fields relax the mean field approximation by allowing atomic charge distributions to adapt to their electrostatic environments.
A broad range of polarizable models have been proposed and reviewed \cite{lemkul_empirical_2016,lopes_molecular_2009, cieplak_polarization_2009,  ponder_force_2003, halgren_polarizable_2001} including
Quantum mechanical methods such as X-POL \cite{xie_design_2007} or full QM-MM \cite{senn_qmmm_2009, lin_qmmm_2007}, fluctuating charge models \cite{rick_dynamical_1994, rick_dynamical_1996, stern_combined_2001, patel_charmm_2004, patel_charmm_2004-1, rappe_charge_1991, senftle_reaxff_2016, khajehpasha_cent2_2022,ghasemi_interatomic_2015,ko_fourth-generation_2021, ko_general-purpose_2021, sprik_polarizable_1988}, and self-consistent point dipole approaches \cite{liu_constructing_1998, kaminski_development_2002, ren_consistent_2002, jorgensen_polarization_2007, ponder_current_2010, shi_polarizable_2013, laury_revised_2015, li_possim_2014}.
In this paper, we focus  on Drude oscillator models, which assign each atom a variable dipole moment but represent them as physical dipoles composed of a positively charged core harmonically attached to a negatively charged shell \cite{lemkul_empirical_2016, kunz_development_2009, van_maaren_molecular_2001}.
In all these approaches, atomic or molecular charge distributions respond to their environment, allowing for  improved modelling of  heterogeneous systems at the cost of computational efficiency.

The responsiveness of atomic charge distributions to changing electric fields complicates the evaluation of inter-atomic interactions by breaking the additive structure of the fixed charge models.
Since each atom's charge distribution changes as a function of its local field, the electric field that it produces must also change.
However, the change in the field produced by each atom will in turn change the local fields of the other atoms further modifying their atomic distributions.
The atomic charge distributions must then be solved self-consistently before evaluating the interactions between the atoms and we can no longer take a simple pair-wise sum over independent interactions.
These force fields are therefore called ``non-additive".
Different polarizable models vary in their approach to updating atomic distributions.
Some force fields, such as the AMOEBA models \cite{ponder_current_2010, shi_polarizable_2013, laury_revised_2015}, update the dipole moments through self-consistent iteration. 
Drude oscillator models are typically updated using an extended Lagrangian method \cite{rick_dynamical_1994, van_belle_calculations_1987, lamoureux_modeling_2003}, inspired by Car-Parinello Molecular Dynamics \cite{car_unified_1985}, that assign the Drude particle a small fictitious mass and then propagates it as an independent dynamical variable.
To mimic self consistent minimization a dual thermostat scheme is employed that sets the center of mass motion of the atoms at the desired temperature of interest while the relative motion of the core and shell is held at a very low temperature, typically 1 K.

While the extended Lagrangian method dramatically accelerates dynamics by eliminating the need for self consistent iteration, it can present new complications.
The addition of a charged, low mass Drude particle can cause numerical instabilities when it approaches another charged atom leading to large forces and catastrophic polarization \cite{lemkul_empirical_2016}.
Furthermore, the extended Lagrangian method can violate equipartition due to the flow of energy from the physical degrees of freedom to the fictitious Drude particle \cite{son_proper_2019}.
Several methods have been proposed for mitigating these issues, including the addition of a hard wall constraint to the Drude dipoles that prevents catastrophic polarization \cite{chowdhary_polarizable_2013}, and the Temperature-Grouped Nosé-Hoover thermostat \cite{son_proper_2019} which decreases the energy flow between the physical and fictitious degrees of freedom.

We will show that several of these issues are exacerbated by an unphysical coupling of the dipole degree of freedom, represented by the displacement between Drude core and shell particles, and non-Coulombic forces.
In section \ref{sec:theory}, we begin by demonstrating how this erroneous coupling arises due to assigning the entirety of the non-Coulombic forces to the Drude core and how it can be corrected by symmetrically assigning these forces.
Section \ref{sec:results} presents numerical examples that illustrate how the force field asymmetry can lead to violations of equipartition and the induction of dipoles in the absence of any electric fields - issues that are resolved by suitably symmetrizing the potential.

\section{Theory and Methods}
\label{sec:theory}

\subsection{The Drude Oscillator Model}
\label{subsec:Drude}

The purpose of polarizable force fields is the construction of a model for the response of atomic charge distributions to an electric field $\bm{E}$.
We will assume that each atom responds linearly to the applied field with a polarizability $\alpha$, so that the atomic dipole $\bm{\mu}$ is given by:
\begin{equation}
    \label{eq:alpha-mu}
    \bm{\mu} = \alpha \bm{E}.
\end{equation}
This linear-response approximation is well-justified in typical condensed phase systems where the electric fields only lead to a modest change in the electronic structure.
Higher order multipole moments are typically neglected although there is no restriction that prevents their inclusion if a more detailed description of short range electrostatics is needed.
In general, atoms may polarize anisotropically (e.g. polarizing more easily in the direction of bonds) in which case $\alpha$ would be represented by a tensor.
For simplicity, we will assume that each atom is also isotropically polarizable so that $\alpha$ is a scalar number.
This does not modify the arguments and concepts presented in this paper but complicates the notation.

The Drude oscillator model represents each atom as a core particle with charge $q_c$ at position $\bm{r_c}$ and a shell particle with charge $q_s$ at position $\bm{r_s}$.
The core and shell charges must sum to the total charge of the atom $q = q_c + q_s$.
Since we will be primarily interested in the dipole degree of freedom we will take $q=0$ noting that since the total atomic charge is taken to be fixed in the Drude oscillator model this contribution can simply be added back in using the fixed point charge potential in Eq. (\ref{eq:UCfp}).
This assumption simplifies our notation and allows us to define the Drude charge $q_D = q_s = -q_c$.

The Drude core and shell are connected by a harmonic potential with spring constant $k_D$ which allows us to relate the Drude model to the polarizable dipole expression in Eq.(\ref{eq:alpha-mu}).
In the absence of an applied field, the minimum core-shell displacement $\bm{d}\equiv \bm{r_c} - \bm{r_s}$, vanishes corresponding to no atomic dipole $\bm{\mu} := q_D\bm{d}=0$. \footnote{Rigorously, we can only relate the idealized point dipole $\bm{\mu}$ to a physical dipole comprised of two point charges if the Drude core and shell are very close. Explicitly, $d$ must be much smaller than the spatial variations in the electric field in any direction $\bm{\hat{r}}$ so that $d \ll \mathrm{max}_{\bm{\hat{r}}}\lbrace \bm{\nabla_{\hat{r}}} \cdot \bm{E}\rbrace/E$. The field exerted by the point and physical dipoles will only agree at distances $r \gg d$. The near field contribution at shorter distances requires higher order multipole terms. }
More generally, when an electric field is applied, the minimum energy displacement will shift, yielding a dipole moment that aligns with the field
\begin{equation}
    \label{eq:mu_min}
    \bm{\mu^*} = q_D\bm{d^*}= \frac{q_D^2}{k_D} \bm{E},
\end{equation}
where we have used the fact that $\bm{d^*}$ is the minimum of the shifted harmonic $k_D/2 |d|^2 -q_d \bm{E}\cdot\bm{d}$.
We can then compare this expression to Eq. (\ref{eq:alpha-mu}) to express the polarizability in terms of Drude oscillator parameters
\begin{equation}
    \label{eq:alpha}
    \alpha = \frac{q_D^2}{k_D}.
\end{equation}

The polarizability $\alpha$ therefore does not fully specify the Drude oscillator parameters $q_D$ and $k_D$ since for any choice of $q_D$ or $k_D$ there exists a choice of the other parameter that yields a desired $\alpha$.
There are two commonly used conventions for selecting these Drude parameters.
One convention, which is used in the CHARMM Drude force field \cite{lemkul_empirical_2016}, is to fix the spring constant to a specified value of $k_D = 1000$ $\mathrm{kCal /mol / \AA ^{2}}$ and to change the charge $q_D\to \sqrt{k_D \alpha}$ to match the polarizability \cite{lamoureux_modeling_2003}.
This approach is appealing since the spring constant $k_D$ plays an important role in determining both the time step and the typical fluctuations of the displacement vector $\bm{d}$ which controls how closely the Drude oscillator represents a point dipole.
Fixing the spring constant therefore makes it easier to control the numerical performance of the model.
Alternatively, the Drude charge can be fixed at $q_D= -1e$ and vary the spring constant $k_D\to q_D^2/\alpha$ \cite{schroder_simulating_2010}.

The Drude oscillator model is non-additive since the electric field in Eq. (\ref{eq:mu_min}) depends on all dipole moments in the system.
The atomic dipoles and  electric field must be solved using a self consistent  field (SCF) approach, requiring an expensive iterative minimization.
However, this SCF minimization can be avoided through an extended Lagrangian scheme that treats the dipoles as dynamical variables.
In the Drude oscillator model this is done by splitting the mass of each atom between the Drude core and shell so that $M_i = m^{(i)}_c + m^{(i)}_D$, where we have used the notation $m_D$ for the shell mass to highlight that this is a key parameter of the Drude algorithm, and reintroduced the atom label $i$.
The centers of mass of the Drude oscillators with coordinates,
\begin{equation}
    \label{eq:RCoM}
    \bm{R_i} = \frac{m_c^{(i)}}{M_i}\bm{r_c^{(i)}} + \frac{m_s^{(i)}}{M_i}\bm{r_s^{(i)}},
\end{equation}
along with the positions of any non-polarizable atoms are evolved with a thermostat at the temperature of interest $T$.
The relative motion of the core-shell pairs,
\begin{equation}
    \label{eq:d}
    \bm{d_i} = \bm{r_c^{(i)}} - \bm{r_s^{(i)}},
\end{equation}
are instead held at a low temperature $T_D$ to keep them close to their energetic minima and mimic the SCF minimization.

\subsection{Spurious Polarization in Drude Force Fields}
\label{subsec:asymmetric}

The Drude oscillator approach presents one more important decision that must be made to specify the model. 
How do we divide the non-Coulombic forces, $U_\mathrm{short}$ and $U_\mathrm{bond}$ in Eq. (\ref{eq:ff_struct}), between the core and shell particles?
In particular, we must ensure that the division of forces between core and shell agrees with the SCF solution of $\lbrace\bm{\mu_i}\rbrace$ in Eq. (\ref{eq:mu_min}), with the dipole moment responding only to the local electric field and not to Coulomb forces.
Thus far, this has been done by assigning the entirety of the non-Coulomb forces to the core particles and none to the shells so that
\begin{equation}
    \label{eq:pot-asym}
    U_\mathrm{short}(\lbrace\bm{r_i}\rbrace) + U_\mathrm{bond}(\lbrace\bm{r_i}\rbrace) \to U_\mathrm{short}(\lbrace\bm{r_c^{(i)}}\rbrace) + U_\mathrm{bond}(\lbrace\bm{r_c^{(i)}}\rbrace),
\end{equation}
where we simply evaluate all non-electrostatic potentials and forces at the position of the Drude core.
The resulting forces on the core and shell are then given by
\begin{subequations}
    \label{eqs:FF-asym}
    \begin{equation}
        \begin{split}
        \bm{F_c^{(i)}} &= -\frac{\partial U_\mathrm{short}(\lbrace\bm{r_c^{(i)}}\rbrace)}{\partial \bm{r_c^{(i)}}}
        -\frac{\partial U_\mathrm{bond}(\lbrace\bm{r_c^{(i)}}\rbrace)}{\partial \bm{r_c^{(i)}}} + (q_D^{(i)} +q_i) \bm{E} - k_D^{(i)}(\bm{r_s^{(i)}} - \bm{r_c^{(i)}})\\
        &:= \bm{F_\mathrm{short}^{(i)}} + \bm{F_\mathrm{bond}^{(i)}} +(q_D^{(i)}+q_i)\bm{E} - k_D^{(i)}(\bm{r_s^{(i)}} - \bm{r_c^{(i)}})
        \end{split}
    \end{equation}
    \begin{equation}
        \bm{F_s^{(i)}} = -q_D^{(i)} \bm{E} + k_D^{(i)}(\bm{r_s^{(i)}} - \bm{r_c^{(i)}})
    \end{equation}
\end{subequations}
where $\bm{F_c^{(i)}}$ and $\bm{F_s^{(i)}}$ are the forces acting on the $i^\mathrm{th}$ core and shell respectively.
We have also reintroduced the contributions of the atomic charges $q_i$ through the fixed point charge model of Eq. (\ref{eq:UCfp}).
This convention is motivated by thinking of the Drude core as a physical degree of freedom representing the nuclear center of mass while the shell is the fictitious degree of freedom introduced by the extended Lagrangian scheme.

The SCF minimization in Eq. (\ref{eq:mu_min}) that we would like to compare to is expressed in terms $\bm{d_i}$ and so we must evaluate the forces acting on the relative degrees of freedom $\lbrace \bm{R_i}, \bm{d_i}\rbrace$.
Transforming Eq. (\ref{eqs:FF-asym}) into the relative coordinate system defined by Eqs. (\ref{eq:RCoM}) and (\ref{eq:d}) yields
\begin{subequations}
    \label{eqs:FF-asym-reduced}
    \begin{align}
        \bm{F_R^{(i)}} = \bm{F_{c}^{(i)}} + \bm{F_s^{(i)}} =  \bm{F_\mathrm{short}^{(i)}} + \bm{F_\mathrm{bond}^{(i)}} + q_i\bm{E}
    \end{align}
    \begin{equation}
        \bm{F_d^{(i)}} = \frac{m_s^{(i)}}{M_i}\bm{F_c^{(i)}} - \frac{m_c^{(i)}}{M_i}\bm{F_s^{(i)}} = \frac{m_s^{(i)}}{M_i}\left(\bm{F_\mathrm{short}^{(i)}} + \bm{F_\mathrm{bond}^{(i)}} +q_i\bm{E}\right) + q_D^{(i)} \bm{E} - k_D^{(i)} \bm{d_i}.
    \end{equation}
\end{subequations}
The resulting force expression should raise concerns since the force acting on the relative motion of the core and shell contains a contribution from non-Coulombic forces.

We can compare this directly to Eq. (\ref{eq:mu_min}) by noting that the minimum arises when $\bm{F_d^{(i)}}=0$, giving
\begin{equation}
    \label{eq:mu_min-asym}
    \bm{\mu_i^{(\mathrm{asym})}}=q_D^{(i)}\bm{d_i^{(\mathrm{asym})}} = \frac{q_D^{(i)}}{k_D^{(i)}}\frac{m_s^{(i)}}{M_i}\left(\bm{F_\mathrm{short}^{(i)}} + \bm{F_\mathrm{bond}^{(i)}} + q_i\bm{E}\right) + \frac{{q_D^{(i)}}^2}{k_D^{(i)}} \bm{E}.
\end{equation}
The first term of this expression clearly differs from the SCF result.
Moreover, it couples the dipole moment to non-Coulombic fields and therefore polarizes atomic dipoles in the absence of any electric field.
The $m_s$ dependence of this error is also surprising since we arrived at Eq. (\ref{eq:mu_min-asym}) using purely energetic arguments that depend only on configurational and not dynamical information.
Traditional Drude force fields therefore display static distributions that depend on mass, a situation that should not arise in thermodynamic equilibrium.
Moreover, these arguments do not rely on the thermostat at all, assuming only that $\lbrace\bm{d_i}\rbrace$ are minimized holding the centers of mass fixed.
The same error should therefore also appear for a SCF minimization using the same force fields and coordinates even though the SCF approach makes no reference to the fictitious masses.

The counter-intuitive mass dependence of the error arises not from any dynamical effects but rather from the mass-dependent relative coordinate  system in Eqs. (\ref{eq:RCoM}) and (\ref{eq:d}).
The extended Lagrangian method is thermodynamically consistent with an exact SCF minimization of the degree of freedom that is held at a low temperature by the dual thermostat,  in this case $\bm{d_i}$.
For any finite shell mass, $m_s$, the response of this coordinate includes contributions from both the core and shell - although the core contribution vanishes in the $m_s\to 0$ limit.
The asymmetric force field in Eq. (\ref{eqs:FF-asym}) assigns too much of the non-Coulomb potential to the core, leading to an excess response of the core than is accounted for by the center of mass.
This excess core response with no concomitant shell response spills over into the relative coordinate, polluting the dipole response.
We see that the spurious force therefore originates from the discrepancy between the division of the potential between the core and shell and the definition of the relative coordinate system.

\begin{figure}
    \centering
    \includegraphics[width=\textwidth]{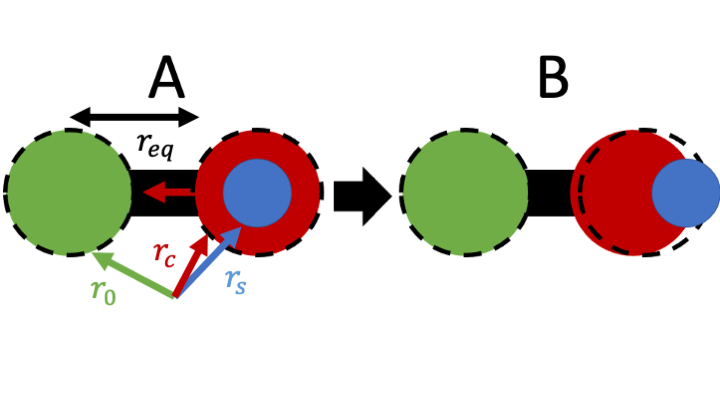}
    \caption{Schematic of a Drude atom comprised of a core (red) and shell (blue) bonded to a non-polarizable atom (green). The  molecule is  shown extended from its equilibrium bond length $r_{eq}$. The centers of mass which are held fixed are shown as dashed circles. (A) The asymmetric bonded force acting on the core is shown as a red arrow with the relative degree of freedom is initially set to $\bm{d}=\bm{0}$. (B) The minimized configuration where the bonded force and Drude spring force are balanced fixing the centers of mass, showing the spurious atomic polarization.}
    \label{fig:asym_bond}
\end{figure}

We can make the origin of these spurious polarizations concrete by considering a simple model molecule comprised of a neutral polarizable atom, treated as a Drude oscillator, bonded to a neutral non-polarizable atom - sketched in Fig. \ref{fig:asym_bond}.A.
In the absence of an electric field, we know that the Drude atom should show no net polarization in every atomic configuration, regardless of the bonded force, i.e. independent of the displacement of the bonded atom from their equilibrium geometry.
We can then perform the self-consistent minimization of the relative degree of freedom under the constraint that the center of mass of all atoms is held fixed - the SCF procedure that corresponds to the dual Drude thermostat.
The force experienced by the Drude core due to its bond to the non-polarizable atom is given by 
\begin{equation}
    \bm{F}_\mathrm{bond} = -k(\bm{r_c} - \bm{r_0})\left(1 - \frac{r_{\mathrm{eq}}}{|\bm{r_c} - \bm{r_0}|}\right),
\end{equation}
where $\bm{r_C}$ and $\bm{r_0}$ are the positions of the Drude core and non-polarizable atom respectively, $k$ is the bond constant and $r_\mathrm{eq}$ is the equilibrium bond length.

Since the core position contributes to the relative Drude coordinate, $\bm{d}$, being minimized, the core can move in response to this force.
However, since the minimization is performed holding the center of masses fixed, the Drude shell must also move in the opposite direction $\sim (\bm{r_C}-\bm{r_0})$ in order to keep the center of mass fixed.
This will then impose a force on the core-shell displacement due to the Drude spring.
The dipole moment corresponding to the energy minimization is reached when these two forces are exactly balanced and is given by Eq. \ref{eq:mu_min-asym} as
\begin{equation}
 \bm{\mu_i^{(\mathrm{asym})}}= -q_D \frac{m_s}{M}\frac{k}{k_D}\left(1-\frac{r_\mathrm{eq}}{|\bm{r_C}-\bm{r_0}|}\right)(\bm{r_C} - \bm{r_0}).
\end{equation}
This expression clearly depends unphysically on the position of a neutral non-polarizable atom through $r_{\mathrm{eq}}$.
Furthermore, we see that a dipole moment has formed even in the absence of any external field.
This arose since both the core and shell are included in the $\bm{d}$ coordinate and the force is not properly split between them.
This leads to the core responding more strongly than the shell which must then move in the opposite direction to maintain the fixed center of mass.
This motion then leads to the formation of a spurious polarization in the absence of an electric field shown in Fig. \ref{fig:asym_bond}.B.
As the shell mass approaches the limit $m_s\to 0$, the shell particle must move more to compensate for a small displacement of the core.
This larger shell displacement would lead to a larger restoring force from the Drude spring limiting how much the core can move to relax the bond force.

\subsection{Symmetrized Drude Force Fields}
\label{subsec:symmetric}

We can find the correct symmetrized division of the non-dipole forces by proposing  the following ansatz:
\begin{subequations}
    \label{eq:ff-ansatz}
    \begin{equation}
        \begin{split}
            U_\mathrm{short}(\lbrace\bm{r_i}\rbrace) + U_\mathrm{bond}(\lbrace\bm{r_i}\rbrace) &\to U^{(c)}_\mathrm{short}(\lbrace\bm{r_c^{(i)}}\rbrace) + U^{(c)}_\mathrm{bond}(\lbrace\bm{r_c^{(i)}}\rbrace)\\ &+ U^{(s)}_\mathrm{short}(\lbrace\bm{r_s^{(i)}}\rbrace) + U^{(s)}_\mathrm{bond}(\lbrace\bm{r_s^{(i)}}\rbrace)
        \end{split}
    \end{equation}
    \begin{equation}
        U^{(c)}_\mathrm{short}(\lbrace\bm{r_c^{(i)}}\rbrace) + U^{(c)}_\mathrm{bond}(\lbrace\bm{r_c^{(i)}}\rbrace)=a_i\left(U_\mathrm{short}(\lbrace\bm{r_c^{(i)}}\rbrace) + U_\mathrm{bond}(\lbrace\bm{r_c^{(i)}}\rbrace)\right)
    \end{equation}
    \begin{equation}
        U^{(s)}_\mathrm{short}(\lbrace\bm{r_s^{(i)}}\rbrace) + U^{(s)}_\mathrm{bond}(\lbrace\bm{r_s^{(i)}}\rbrace)=b_i\left(U_\mathrm{short}(\lbrace\bm{r_s^{(i)}}\rbrace) + U_\mathrm{bond}(\lbrace\bm{r_s^{(i)}}\rbrace)\right)
    \end{equation}
    \begin{equation}
        q_c^{(i)} = a_i q_i + q_D^{(i)}
    \end{equation}
    \begin{equation}
        q_s^{(i)} = b_i q_i - q_D^{(i)},
    \end{equation}
\end{subequations}
where $a_i$ and  $b_i$ are coefficients that define how the force field is divided between the particles.
Determining an appropriate choice of these coefficients that eliminates the spurious polarization will be our primary goal in this section.

Before proceeding, there are several features of Eq. (\ref{eq:ff-ansatz}) that should be highlighted.
First, we note that, in contrast  to traditional Drude force fields, we have divided the bare atomic charge between the core and the shell rather than placing the entirety of the charge on the core.
This is needed so that the dipole response is not polluted by the monopole response, the $q\bm{E}$ term in Eq.  (\ref{eq:mu_min-asym}).
Second, there is some subtlety to where the forces are evaluated in the additive and polarizable force fields.
In the additive force fields, forces and potentials are evaluated at the atomic center of mass while in the Drude force field they are evaluated at the  positions of the core and shell.
Strictly, the forces and potentials should in all cases be evaluated at the center of mass but then assigned to the cores and shells weighted by $\alpha$ and $\beta$.
However, this can introduce substantial challenges in implementing these force fields in traditional molecular dynamics software.
Instead, we can retain the feature of traditional Drude force fields that treat the  core and shell just like any other atom in the system evaluating the forces at their positions.
This approach is an approximation and can lead to residual asymmetry in the force field due to differences in the potentials evaluated at the positions of the core and shell, which we refer to as localization error.
As a result, some of the spurious polarization in Eq. (\ref{eq:mu_min-asym}) may remain due to this error.
We will quantify the localization error in more depth below in section \ref{sec:loc_err}, but we note that it is negligible when $d$ is much smaller than the length scale of variations in the forces.
This can be controlled by increasing the value of the Drude spring constant $k_D$ or eliminated entirely by evaluating the forces at the Centers of mass.
In either case, we will drop the explicit coordinate dependence  and assume that $a^{-1}(U^{(c)}_\mathrm{short}+ U^{(c)}_\mathrm{bond})\approx U_\mathrm{short} + U_\mathrm{bond}\approx b^{-1}(U^{(s)}_\mathrm{short} + U^{(c)}_\mathrm{bond})$.

With these caveats in place, we can now proceed by evaluating the forces acting on the relative degrees of freedom, giving the following expressions:
\begin{subequations}
    \label{eq:ff-ansatz-relative}
    \begin{align}
        \bm{F_R^{(i)}} =  (a_i+b_i)\left(\bm{F_\mathrm{short}^{(i)}} + \bm{F_\mathrm{bond}^{(i)}} + q\bm{E}\right)
    \end{align}
    \begin{equation}
        \bm{F_d^{(i)}} =\left(a_i\frac{m_s^{(i)}}{M_i} - b_i \frac{m_c^{(i)}}{M_i}\right)\left(\bm{F_\mathrm{short}^{(i)}} + \bm{F_\mathrm{bond}^{(i)}} +q\bm{E}\right) + q_D \bm{E} - k_D \bm{d_i}.
    \end{equation}
\end{subequations}
Requiring the center of mass forces to be consistent with the force on the atoms in the additive force field and the force on the relative degree of freedom to vanish at the SCF minimum in Eq. (\ref{eq:mu_min}) yields a  simple system of equations for $a_i$ and $b_i$,
\begin{subequations}
    \label{eq:sym-condition}
    \begin{equation}
        a_i + b_i =1
    \end{equation}
    \begin{equation}
        a_i\frac{m_s^{(i)}}{M_i} = b_i \frac{m_c^{(i)}}{M_i}
    \end{equation}
    \begin{equation}
        a_i =\frac{m_c^{(i)}}{M_i}
    \end{equation}
    \begin{equation}
        b_i =\frac{m_s^{(i)}}{M_i}.
    \end{equation}
\end{subequations}
The symmetrization result has a simple interpretation.
The non-Coulomb forces and atomic charge must be divided between the core and the shell weighted by their mass.
This procedure yields a set of relative  coordinates and forces that are consistent with the SCF procedure and avoid the spurious polarizations that appear in the standard, asymmetric Drude force fields, so that Eq. \ref{eq:ff-ansatz-relative} reduces to
\begin{subequations}
    \label{eq:ff-sym}
    \begin{align}
        \bm{F_R^{(i)}} =  \left(\bm{F_\mathrm{short}^{(i)}} + \bm{F_\mathrm{bond}^{(i)}} + q\bm{E}\right)
    \end{align}
    \begin{equation}
        \bm{F_d^{(i)}} = q_D \bm{E} - k_D \bm{d_i}.
    \end{equation}
\end{subequations}
The SCF minimization of $\bm{d_i}$ is then given, as above, by setting $\bm{F_d^{(i)}}=\bm{0}$ with resulting dipole moments
\begin{equation}
    \label{eq:mu_min-sym}
    \bm{\mu_i^{(\mathrm{sym})}}=q_D^{(i)}\bm{d_i^{(\mathrm{sym})}} = \frac{{q_D^{(i)}}^2}{k_D^{(i)}} \bm{E},
\end{equation}
in perfect agreement with the desired result in Eq. (\ref{eq:mu_min}).

\subsection{Symmetrized Pairwise Potentials}
\label{sec:sym-pair-pots}

The symmetrization conditions in Eq. (\ref{eq:sym-condition}) express the relationship between the potentials experienced by cores and shells and can be satisfied by many different force fields.
In this section, we will show how we can concretely construct pairwise potentials that can be applied in molecular simulations.
These pairwise specifications will apply to both non-bonded pair potentials and bond potentials and will illustrate how they can be generalized to three and four body potentials such as angle, proper and improper dihedral, and non-bonded Stillinger-Weber potentials - though the algebra will be more involved and may incur additional computational cost in implementation.

The central question in symmetrizing pairwise potentials is how we can apply the specification in Eq. (\ref{eq:sym-condition}) to potentials of the form
\begin{equation}
\begin{split}
    \label{eq:ff-pairwise}
    U\left(\left\lbrace\bm{r_c^{(i)}}, \bm{r_s^{(i)}}, \bm{r_i}\right\rbrace\right) = \frac{1}{2}\left(\sum_{i\neq j}u_{cc}^{(i,j)}(\bm{r_c^{(i)}}-\bm{r_c^{(j)}}) + \sum_{i\neq j}u_{ss}^{(i,j)}(\bm{r_s^{(i)}}-\bm{r_s^{(j)}})\right) \\ +\sum_{i\neq j}u_{cs}^{(i,j)}(\bm{r_c^{(i)}}-\bm{r_s^{(j)}}) + \frac{1}{2}\sum_{i\neq j}u_{nn}^{(i,j)}(\bm{r_i}-\bm{r_j})\\
    + \sum_{i, j}u_{cn}^{(i,j)}(\bm{r_c^{(i)}}-\bm{r_j}) + \sum_{i, j}u_{sn}^{(i,j)}(\bm{r_s^{(i)}}-\bm{r_j}),
\end{split}
\end{equation}
where $\bm{r_c^{(i)}}$ and $\bm{r_s^{(i)}}$ are the core and shell positions of the i$^{th}$ polarizable atom and $\bm{r_i}$ is the position of the i$^{th}$ non-polarizable atoms.
The polarizable and non-polarizable atoms are distinct lists so that $\bm{r_i}$ and $\bm{r_c^{(i)}}$ refer to different atoms.
Equation (\ref{eq:ff-pairwise}) contains core-core, core-shell and shell-shell potentials as well as interactions between cores and shells and non-polarizable particles which can all generally be different.
Throughout, we will assume that evaluating any of the potential functions at the core and shell position of a given atom will yield the same answer.
This is the same as the small displacement assumption we applied to derive the symmetrized force fields.

We are interested in constructing symmetrized force fields that start from a non-polarizable or asymmetric force field $u_{ij}$ that has already been parameterized.
Our goal is then to determine how this force field can be split between  the cores and shells, motivating the ansatz
\begin{equation}
    u_{\alpha\beta}^{(i,j)} = a_{\alpha\beta}^{(i,j)} u_{ij},
\end{equation}
where $\alpha,\beta \in\lbrace c,s,n\rbrace$ label whether the atoms are cores, shells or non-polarizable and $i, j$ are atom indexes.
This ansatz is particularly convenient since many pair potentials include a scalar parameter that sets their energy scale (e.g. the bond spring constant or the $\epsilon$ parameter  in Lennard-Jones and WCA potentials).
Under this scaling ansatz, symmetrization can then be accomplished by simply rescaling these force field parameters.
Since, we seek a prescription for symmetrizing force fields that works for any number of particles in the system, the potentials should be symmetrized for each pair of atoms and Equation (\ref{eq:sym-condition}) should therefore apply to each pair of atoms.

Consider first the pair potentials involving a non-polarizable particle.
For interactions between, two non-polarizable particles $a_{nn}^{(i,j)}=1$ since there are no cores or shells to split the interaction between.
Similarly, we can immediately use Eq. (\ref{eq:sym-condition}) to split the interactions between non-interacting particles and cores or shells to give 
\begin{subequations}
    \label{eq:sym-FF-NP}
    \begin{equation}
       a_{cn}^{(i,j)}=m_c^{(i)}/M_i 
    \end{equation}
    \begin{equation}
        a_{sn}^{(i,j)}=m_s^{(i)}/M_i.
    \end{equation}
\end{subequations}
This solution is unique since we have  two constraints for two variables.

The  situation is more complicated for interactions between two Drude pairs.
This leaves us with four potentially independent parameters that govern the four possible interactions between the cores and shells on different atoms, $a_{cc}^{(i,j)}$, $a_{ss}^{(i,j)}$, $a_{cs}^{(i,j)}$, and $a_{cs}^{(j,i)}$.
This gives us a system of four linear equations for the coefficients
\begin{subequations}
    \begin{equation}
        a_{cc}^{(i,j)} + a_{cs}^{(i, j)} = \frac{m_c^{(i)}}{M_i}
    \end{equation}
    \begin{equation}
        a_{cc}^{(i,j)} + a_{cs}^{(j, i)} = \frac{m_c^{(j)}}{M_j}
    \end{equation}
    \begin{equation}
        a_{cs}^{(j,i)} + a_{ss}^{(i, j)} = \frac{m_s^{(i)}}{M_i}
    \end{equation}
    \begin{equation}
        a_{cs}^{(i,j)} + a_{ss}^{(i, j)} = \frac{m_s^{(j)}}{M_j}.
    \end{equation}
\end{subequations}
While this would at first appear to uniquely specify the four pair potential specifications, only three of these equations are linearly independent.
This indicates that we will have a one parameter family of possible force field symmetrizations, or alternatively that we are free to impose one additional independent constraint on the coefficients.

We can take advantage of this flexibility by imposing the condition that $a_{ss}^{(i,j)}=0$ in order to decrease the number of pair force evaluations we need to perform while simultaneously minimizing the force on the lowest mass shell particles that are most prone to numerical error.
This yields the following prescription for the symmetrized force fields between Drude atom pairs
\begin{subequations}
    \label{eq:sym-FF-cs}
    \begin{equation}
        a_{cc}^{(i,j)} = \frac{m_c^{(i)}}{M_i} + \frac{m_c^{(j)}}{M_j} - 1
    \end{equation}
    \begin{equation}
        a_{cs}^{(i, j)} = \frac{m_s^{(j)}}{M_j}
    \end{equation}
    \begin{equation}
        a_{ss}^{(i, j)} = 0.
    \end{equation}
\end{subequations}

In general, if we  are interested in symmetrizing an $m$ body potential involving $n$ Drude oscillator atoms, we must solve a system of $2^{n}$ linear equations for the same number of coefficients.
For example,  a three body interaction involving three  Drude oscillator pairs would have 8 coefficients $u_{\alpha\beta\delta}^{(i,j,k)}$, where $\alpha,\beta,\delta=c$ or $s$.
This is no more  complicated than the pair potentials we discuss here but becomes algebraically more difficult.
In addition, it  can rapidly increase the number of potentials that must be ccomputed increasing computational complexity.
In general, several of these constraints may be  linearly dependent allowing us to impose further  constraints that can reduce the number of potential terms that need to be evaluated. 

It is possible to only symmetrize some non-Coulombic terms in the force field since they act additively (e.g. we simply add together the bond and angle potentials).
This partial symmetrization will alleviate the spurious polarizations that arise from the symmetrized terms but will retain the errors from unsymmetrized potentials.
Such an approach may be desirable when the cost of reparameterizing a potential or computing the additional terms required by symmetrization is too high relative to the mitigation in error.
In general, the most important degrees of motion to symmetrize are those that are similar in frequency to the Drude oscillator since these will be the ones that can transfer energy most efficiently to the unphysical degrees of freedom.
For most systems of interest using the constant $k_D$ convention, this will be the bond vibrations which can vibrate close to resonance with Drude oscillators \cite{son_proper_2019}, and non-bonded collisions which lead to impulsive $\delta$ function like forces that contain a broad range of frequencies and can be very large in magnitude.
Fortunately, these are typically pairwise potentials and can be symmetrized as described in this section with modest computational cost.

\subsection{Localization Error}
\label{sec:loc_err}

After symmetrizing the potential as described above, the primary remaining source of error is the error from evaluating the forces at the core and shell positions instead of at the center of mass.
In this section, we quantify this localization error by examining how the force acting on the centers of mass, and relative degrees of freedom differ when the force is evaluated at the core and shell positions compared to evaluating them at the  center of  mass.
Throughout, we  will assume that the displacement parameter is small relative to the spatial variations in the force and can therefore be used as  a small parameter in a Taylor expansion  to  estimate the error.

We begin by considering the forces acting on the center of  mass and relative Drude coordinates when the asymmetric  force field is evaluated at the core and shell positions,
\begin{subequations}
    \begin{equation}
        \label{eq:FR-loc-asym}
        \bm{F_R}^{(i,asym)} =  \left(\bm{F_\mathrm{short}}^{(i)} + \bm{F_\mathrm{bond}}^{(i)} + q\bm{E}\right)_{\bm{R_i}-\frac{m_s^{(i)}}{M_i}\bm{d_i}}
        + q_D\left(\bm{E}|_{\bm{R_i}-\frac{m_s^{(i)}}{M_i}\bm{d_i}}-\bm{E}|_{\bm{R}+\frac{m_c^{(i)}}{M_i}\bm{d_i}}\right) + k_D \bm{d_i}
    \end{equation}
    \begin{equation}
        \label{eq:Fd-loc-asym}
        \bm{F_d}^{(i,asym)} =  -\frac{m_s^{(i)}}{M_i}\left(\bm{F_\mathrm{short}}^{(i)} + \bm{F_\mathrm{bond}}^{(i)} + q\bm{E}\right)_{\bm{R}-\frac{m_s^{(i)}}{M_i}\bm{d_i}} - q_D\left(\frac{m_s^{(i)}}{M_i}\bm{E}|_{\bm{R}-\frac{m_s^{(i)}\bm{d_i}}{M_i}}+\frac{m_c^{(i)}}{M_i}\bm{E}|_{\bm{R}+\frac{m_c^{(i)}}{M_i}\bm{d_i}}\right)  -k_D \bm{d_i},
    \end{equation}
\end{subequations}
where subscripted coordinates indicate where forces and fields must  be evaluated.
Taylor expanding this expression to first order in the Drude displacement $\bm{d_i}$ allows us to estimate the leading order localization error
\begin{subequations}
    \begin{equation}
        \label{eq:FR-loc-err-asym}
        \bm{F_R}^{(i,asym)} \approx  \bm{F_0}|_{\bm{R_i}} -q_D\nabla E|_{\bm{R_i}}\bm{d_i} - \frac{m_D^{(i)}}{M_i}\nabla F_0|_{\bm{R_i}}\bm{d_i} +\mathcal{O}(d_i^2)
    \end{equation}
    \begin{equation}
        \label{eq:Fd-loc-err-asym}
        \bm{F_d}^{(i,asym)} \approx - q_D \bm{E}|_{\bm{R_i}} -k_D \bm{d_i}-\frac{m_s^{(i)}}{M_i}\bm{F_0}|_{\bm{R_i}} - \left(\frac{m_s^{(i)}}{M_i}\right)^2 \nabla F_0|_{\bm{R_i}}\bm{d_i} - \left(\frac{m_c^{(i)}}{M_i}-\frac{m_s^{(i)}}{M_i}\right)\nabla E|_{\bm{R_i}}\bm{d_i} +\mathcal{O}(d_i^2),
    \end{equation}
\end{subequations}
where the non-Drude forces have been collected into the term $\bm{F_0}=\bm{F_\mathrm{short}}^{(i)} + \bm{F_\mathrm{bond}}^{(i)} + q\bm{E}$ for brevity and $\nabla$ is the derivative operator, which takes the form of a matrix for vector valued functions $\bm{F_0}$ and $\bm{E}$.

The first two terms in the center of mass  force, Eq. (\ref{eq:FR-loc-err-asym}), give the expected force if the  Drude  pair were a point dipole located at the center of  mass.
The first term is the non-Drude forces experienced by the center of mass,  while the second is the Coulomb force experiences by the induced dipole $\bm{\mu_i}=q_D\bm{d_i}$.
The third term, however, is a localization error that arises due to evaluating the non-Drude forces at the core position instead of the center of  mass. 
Consequently, we see that Drude force fields can also induce erroneous forces directly on the centers of mass.
The first two terms in the relative force, Eq. (\ref{eq:Fd-loc-err-asym}), also  give the expected force  for a point dipole at the center of mass.
The third term is the asymmetry error discussed above, while the fourth  and fifth terms are new localization errors.
The term proportional to $\nabla F_0$ is the localization error in evaluating the non-Drude forces  while the final term is the localization error in evaluating the electric field.

We repeat this process to estimate the leading order localization error for the symmetrized force field, yielding
\begin{subequations}
    \begin{equation}
        \label{eq:FR-loc-err-sym}
        \bm{F_R}^{(i,sym)} \approx  \bm{F_0}|_{\bm{R_i}} -q_D\nabla E|_{\bm{R_i}}\bm{d_i} +\mathcal{O}(d_i^2)
    \end{equation}
    \begin{equation}
        \label{eq:Fd-loc-err-sym}
        \bm{F_d}^{(i,sym)} \approx - q_D \bm{E}|_{\bm{R_i}} -k_D \bm{d_i}- \frac{m_s^{(i)}m_c^{(i)}}{M_i^2} \nabla F_0|_{\bm{R_i}}\bm{d_i}- \left(\frac{m_c^{(i)}}{M_i}-\frac{m_s^{(i)}}{M_i}\right)\nabla E|_{\bm{R_i}}\bm{d_i} +\mathcal{O}(d_i^2).
    \end{equation}
\end{subequations}
Notably, we see that symmetrizing the force field also eliminates the leading order localization error in the center of mass forces.
The localization error in the relative  degree of freedom of the symmetrized force field is of the same order in $d_i$ as the asymmetric one.
However, the prefactor of the force localization error vanishes more slowly as $m_s^{(i)}\to 0$ in the symmetrized force field compared to the asymmetric force field.
This may yield larger localization error for the symmetrized force field in some systems when the Drude mass is taken to be small.
However, this is likely to be outweighed by the elimination of the asymmetry error which is a lower order error term independent of  $d_i$.

The localization error can be avoided entirely by evaluating the forces at the center of mass and then assigning them to the core and shell according to the symmetrization rules of the previous sections.
The only forces that need to be evaluated at the core and shell  positions are the Coulomb fields acting on the Drude charges  $q_D$ since  these are responsible for the induced dipole forces acting on the polarizable atoms.
Such an approach may be straightforward to implement with minimal computational in some Molecular Dynamics software where the centers of mass of the Drude atoms are already calculated over the course of the time step, and would be advantageous in eliminating all sources of localization  error except for the field localization error acting  on the relative degrees of freedom.

\section{Benchmark Results}
\label{sec:results}

In this section, we demonstrate how applying symmetrized force fields can circumvent some of the numerical challenges that arise in Drude simulations.
A variety of systems will be simulated in the LAMMPS molecular dynamics software \cite{thompson_lammps_2022} using the DRUDE package for thermalized Drude oscillators \cite{dequidt_thermalized_2016}.
This simulation package does not include any corrections for catastrophic polarizations and directly evaluates the equations of motion under the potentials described above.
First, we demonstrate how an asymmetric bond potential in a model polarizable diatomic molecule as shown in Fig. \ref{fig:asym_bond} can lead to energy leakage out of the bond vibration in Section \ref{sec:leakage}, while a symmetrized bond potential avoids this leakage.
When considering a gas of such diatomic molecules in section \ref{sec:flying-ice-cube} we show how the energy leakage leads to violations of equipartition which is attenuated by symmetrizing the potential.
Throughout, we will use parameters that are expected to be the most challenging for Drude oscillators, including resonances between the center of mass degrees of freedom and Drude oscillators as well as large Drude masses, in order to test the force field symmetrization under difficult conditions.

\subsection{Energy Leakage in a Diatomic Molecule}
\label{sec:leakage}

Consider a diatomic molecule comprised of  one non-polarizable atom and one polarizable atom represented by  a Drude oscillator, as drawn in Fig. \ref{fig:asym_bond}.A.
This is the simplest possible system we can consider that differentiates between  the asymmetric and symmetrized force fields.
In order to best highlight the numerical challenges that can arise in Drude oscillator simulations, we will consider the most difficult scenario where $m_c=m_s=m_{np}$ and $k_\mathrm{bond} = k_D$. In this parameter regime, the bond vibrations are resonant with the Drude oscillator and can therefore efficiently transfer energy to the fictitious degree of freedom - a scenario known to cause numerical artefacts \cite{son_proper_2019}.
In the simulations that follow, $k=500$ $\mathrm{kCal \cdot mol^{-1}\cdot\AA^{-2}}$ , $m=1$ amu and  the equilibrium bond distance is $d_\mathrm{bond}^{(eq)}=1\,\AA$.

The initial state is obtained by sampling a bond displacement $d_\mathrm{bond}$ from an equilibrium distribution at $T=300$ K.
It is then propagated with a timestep $dt =1$ fs under a Langevin thermostat for the relative  degree of freedom $\bm{d_\mathrm{Drude}}$ at temperature $T_D=1$ K and damping time $\tau_D=20$ fs, while the center of  mass was not thermostatted. 
This was accomplished  by setting $T_\mathrm{CoM}=300$ K and $\tau_\mathrm{CoM}\to 10^{12}$ fs, much longer than the simulation time  so that the dual Langevin thermostat negligibly impacted the Center of Mass dynamics.

\begin{figure}
    \centering
    \includegraphics[width=\textwidth]{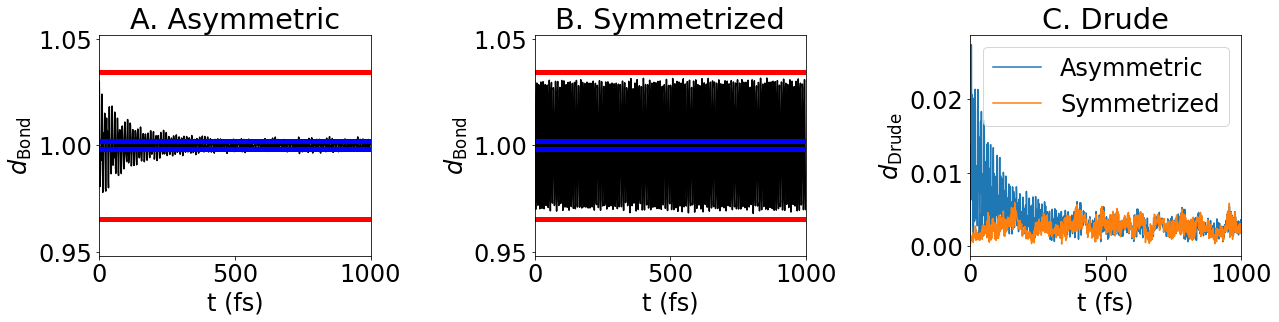}
    \caption{Energy leakage in diatomic molecules with a Drude polarizable atom. Bond length of (A) Asymmetric and (B) Symmetrized force fields. Both plots were initialized in the same initial condition sampled from a Boltzmann distribution at $T=300$ K. Standard deviations at 1K (blue) and 300 K (red) are provided as guides. (C) The Drude displacement coordinate for each of the force  fields is plotted. In all cases, the relative core shell coordinate is held at 1K by a Langevin thermostat, while the centers of mass are not thermostatted.}
    \label{fig:leakage}
\end{figure}

Sample trajectories for  each of the two force fields is shown in Fig. \ref{fig:leakage}, shows  the consequences  of the unphysical coupling between the center of mass and Drude degrees of freedom.
In the asymmetric case shown in Fig. \ref{fig:leakage}.A, this spurious  coupling leads to a rapid, unphysical, loss of energy out of the bond degrees of freedom, damping the vibrations within $\sim 100$ fs.
After the first few 100 fs, the bond vibrations appear to be thermostatted at the Drude temperature, $T_D=1$ K.
In contrast, the symmetrized force field does not show any significant leakage of energy out of the bond vibration.
The symmetrized force field was also propagated for $10^9$ fs, much longer than is shown but still much shorter than the thermostat damping time, and still showed no visible leakage of energy.
The distinction between the force fields is also visible in the relative degree of freedom in Fig. \ref{fig:leakage}, where we see that under asymmetric force fields, dipole moments do not vanish as expected until the bond vibrations have fully cooled.

The particular parameters selected in this example are particularly challenging for Drude oscillator simulations since the bond vibration and Drude oscillator degree of freedom are near resonance.
In addition, since the  mass has been evenly split between the core and shell particle, the error in the asymmetric force field is maximized.
As a result, in many systems the energy leakage into the  Drude oscillator modes is likely less extreme.
Nevertheless, even in  this extreme regime, we see that the symmetrized force field shows an excellent decoupling between the center of mass and dipole dynamics.
In contrast to the asymmetric force field which requires us to select as low a Drude shell mass as numerically feasible, the symmetrized force field continues  to perform well even when the mass is evenly  split between the core and shell.

\begin{figure}
    \centering
    \includegraphics[width=\textwidth]{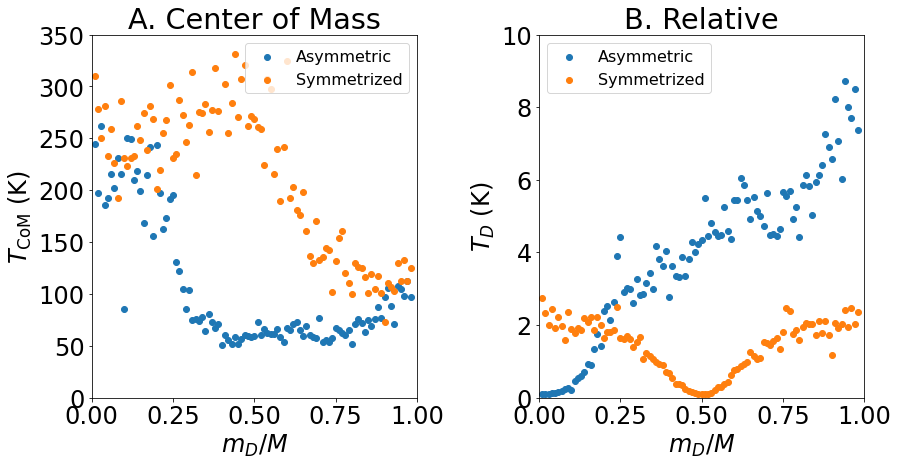}
    \caption{Temperature of (A) Center of Mass coordinate $\bm{R}$ and (B) Relative Drude coordinate $\bm{d}$ as a function of Drude mass. In all cases, a dual thermostat scheme was employed that held the Centers of Mass at $T_\mathrm{CoM}=300$ K and the relative degrees of freedom at $T_D=1$ K.}
    \label{fig:mD_Dep}
\end{figure}

We can further evaluate the performance of the symmetrized force field by introducing a dual thermostat and varying the Drude mass $m_D$, as shown in Fig. \ref{fig:mD_Dep}.
The centers of mass are held at a temperature of $T_\mathrm{CoM}=300$ K, with a damping time of $100$ fs.
The  timestep was scaled by $\sqrt{\min\lbrace\Tilde{m}\rbrace}$, where $\min\lbrace\Tilde{m}\rbrace$ is the smallest reduced mass in the system to ensure that  the integrator stability was conserved as the masses where varied.
Regardless of timestep, the systems were simulated for $100$ ps, discarding the first $1$ ps to  equilibration.
The remaining parameters are the same  as the previous case.
We see that neither the asymmetric nor symmetrized force field are able to fully decouple the Center of Mass and relative motion at all masses.
However,  the  symmetrized force field is able to better maintain the center of mass temperature for a much wider range of Drude masses.
As expected, both force fields describe the centers of mass effectively as $m_D\to0$.
The situation is slightly more complicated for the  relative degree of freedom, where the asymmetric force field maintains the Drude temperature better than the symmetrized force field fir small masses  before turning over as $m_D$ increases.
We  attribute this performance difference due to the more favorable scaling of the localization error in the asymmetric force field as $m_D \to 0$.
Nevertheless, the  symmetrized  force field continues to perform well even in  this scenario.

The improved decoupling allows us significantly more flexibility in selecting larger Drude masses that are more numerically stable and can thereby allow us to  extend the timestep under which Drude oscillator simulations are numerically stable.
This flexibility is important since the Drude oscillator frequency is typically the limiting factor in extending the integration timestep, requiring $\Delta t \sqrt{k/\overline{m}}<1/4\pi$, where $\overline{m}$ is the smallest reduced mass among Drude oscillators.
Increasing the Drude mass for which the force  field remains accurate allows us to extend the timestep from the 1 fs timesteps typically employed in Drude simulations.

Moreover, the flexibility in selecting the mass allows us to tune the  frequency of the Drude oscillator more easily without impacting its polarizability.
As a result, we can select the mass to  avoid resonances with other modes in the system, mitigating energy leakage into the Drude degrees of freedom.
Generally, it is best  to select $\overline{m}$ so that  $|\omega-\sqrt{k/m}| \gg (\gamma + \gamma_D)$ for all modes $\omega$ present in the system, where $\gamma$ and $\gamma_D$ are the friction coefficients of the two thermostats, in order to avoid  this energy leakage.
A recent study has taken the  opposite approach and tuned the mass of Drude particles to match plasmon modes at graphitic-water interfaces \cite{bui_classical_2023} intentionally bringing the Drude oscillators into  resonance with librational modes of water.
This introduces a substantial drag force on the motions of water near the interface, consistent with observed "quantum friction" forces \cite{volokitin_quantum_2011}.
In either case, the flexibility to tune Drude oscillator frequencies without introducing erroneous couplings can enable further optimizations of Drude oscillator simulations.

\subsection{Equipartition Violation in a Gas of Diatomic Molecules}
\label{sec:flying-ice-cube}

Next, we consider the  consequences of this energy leakage on the partitioning of energies in the system.
Previous studies have shown that the transfer of  energy into the Drude oscillators can lead to substantial violations of the equipartition theorem when a velocity rescaling thermostat such as Nos\'e-Hoover is used \cite{son_proper_2019}.
In extreme cases, this leaves Drude oscillator simulations prone to the flying ice-cube artifact where the relative motion of atomic centers of mass are frozen out and the  entire system translates collectively to compensate for this loss of  kinetic  energy.
We now consider  a gas comprised  of 125 of the diatomic molecules simulated above  that can interact through a Lennard-Jones interaction with parameters $\epsilon=0.1$ kCal/mol and  $\sigma=3$ $\AA$.
The Lennard-Jones interactions are also symmetrized in the symmetrized force field but act only between cores in the asymmetric force field.
To simplify our consideration, we do not include Coulomb interactions between dipoles in the system.
The system is propagated with a timestep $\Delta t=1$ fs under a dual Nos\'e-Hoover  thermostat with  $T_{CoM}=300$ K, $T_D=1$ K, $\tau_{CoM}=100$ fs, and $\tau_D=20$ fs.

\begin{figure}
    \centering
    \includegraphics[width=\textwidth]{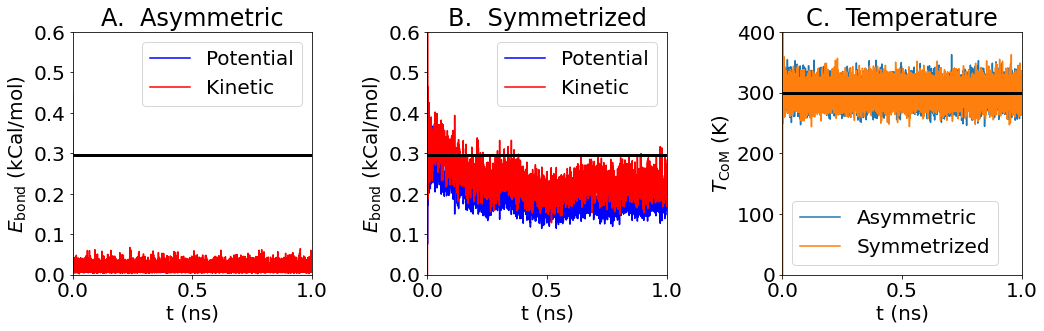}
    \caption{Symmetrized force fields significantly mitigate equipartition violations with Dual Nos\'e-Hoover thermostats in gas of diatomic molecules with one Drude polarizable atom. Bond energies for (A) Asymmetric and (B) Symmetrized force fields. Potential (Blue) and Kinetic (red) Energies are compared to their expected equipartition result (Black). Despite equipartition violations, Temperature (C) of  the Center of Mass degrees of freedom are properly maintained  at the target value of $T=300$ K (Black) for both the asymmetric (blue) and symmetrized (orange) force fields.}
    \label{fig:equipartition}
\end{figure}

Figure \ref{fig:equipartition}.A and B show the kinetic and potential energy in the vibrational degrees of freedom for the asymmetric and symmetrized force fields respectively and compares them with the result expected from the equipartition theorem $\langle E_\mathrm{vib}\rangle =k_B T/2$.
The asymmetric force field shows an extreme deviation from equipartition, and is almost completely frozen out in these dynamics.
In contrast, the error in the symmetrized force fields is much more modest, showing a deviation of  only $10\%$ from the equipartition result even in the extreme parameter regime considered here.
This equipartition violation for the symmetrized force fields is surprising since the isolated molecule showed no energy leakage even when propagated for three orders of magnitude longer.
For the symmetrized force field, some erroneous coupling between the centers of mass and Drude oscillators can arise due to the error incurred by evaluating the Lennard-Jones forces at the position of the core and shell rather than the center of mass.
This error is maximized when the mass is equally split between the core and shell and can couple the Drude oscillator with the bond vibrations with which it is resonant.
It is much smaller for other parameter choices, with the symmetrized force field showing excellent agreement with equipartition.
Figure \ref{fig:equipartition}.C shows that the overall temperature of the centers of mass is properly fixed by the dual thermostat for both force fields considered.
Maintaining the temperature while the vibrations are colder than equipartition indicates that the other, translational and rotational, degrees of freedom in the system must have too much kinetic energy.
The energy leakage in the system therefore introduced error in degrees of freedom that are not directly coupled to the Drude oscillators.

In these dimer gases, equipartition is violated due to the energy leakage out of the bond vibrations discussed in section \ref{sec:leakage}.
The energy in each of the molecular vibrations is rapidly damped out by the spurious coupling to the low temperature Drude thermostat.
The total kinetic energy in the system will then be too low in the system and the velocity rescaling thermostat will increase the velocities in the system to compensate, uniformly increasing the kinetic energy of all degrees of freedom in the system.
The energy leakage into the Drude thermostat and heating up by velocity rescaling eventually balance out to maintain the Center of Mass degrees of freedom at the desired temperature.
However, since energy is leaking out of the vibrational degrees of freedom much more quickly, they will be artificially cooled and the other degrees of freedom must compensate by having an inflated temperature.

\section{Conclusion}
\label{sec:conc}

We have demonstrated that placing all non-Coulomb interactions on the core of a Drude oscillator leads to a potential that spuriously couples atomic polarization to non-Coulombic forces, and which depends explicitly on the masses of the particles in the system.
As a result, both self-consistent minimization and extended Lagrangian methods can yield atomic dipoles in the absence of electric fields.
In addition, when a dual-Thermostat is used that holds the Drude oscillator at a low temperature, energy can leak rapidly out of the center of mass degrees of freedom into the Drude oscillators where it is rapidly quenched.
Significant violations of equipartition can arise due to this energy leakage, polluting the dynamics of degrees of freedom that seem unrelated to the Drude oscillators.

Symmetrizing the non-Drude potentials can significantly mitigate these errors and allows for far more flexibility in selecting the Drude mass.
In particular, symmetrizing the force field yields a potential which formally decouples center of mass and Drude oscillator degrees of freedom affording much more flexibility in the choice of Drude particle mass.
Using traditional asymmetric force fields, the Drude mass must be set as low as computationally feasible to mitigate spurious coupling between the center of mass and Drude oscillator motion and to minimize the mass dependence of the resulting potential.
In contrast, the Drude mass in the symmetrized force fields can be quite large and chosen to improve numerical stability of the simulation, extending the allowable time steps, or to avoid resonances between the Drude oscillators and center of mass motions in the system.
Anecdotally, the cost of symmetrizing the force fields is minimal, with no change in simulation times using the LAMMPS package.
However, symmetrized force fields generally require more force evaluations per time step and will therefore typically incur additional computational cost.
In the simulations presented here, this was outweighed by the cost of other parts of the simulation.

These errors can be challenging to diagnose in simulations for several reasons but can nevertheless be significant.
The magnitude of the spurious force decreases as the shell mass decreases.
In typical simulations, the shell mass is much smaller than the total atomic mass leading to a small prefactor.
For example, in the typical SWM4-NDP model $m_s=0.4 \;\mathrm{g/mol}$ while the total mass of each rigid water is $M= 18.0\;\mathrm{ g/mol}$ \cite{lamoureux_polarizable_2006}, resulting in a $\sim98\%$ attenuation of the spurious force.
However, the underlying forces, particularly $\bm{F_\mathrm{short}}$, can vary significantly in magnitude.
Some configurations, such as during an atomic collision, will have extremely large values of these forces that can significantly impact the dipole moment in those configurations.
During these collisions, the risk of catastrophic polarization is particularly high and may lead to numerical instabilities.
These brief periods of large spurious force can be masked by the application of the hard wall polarization constraint \cite{chowdhary_polarizable_2013} as well as by the strong thermostats typically applied to the dipole degrees of freedom which rapidly damp the effect of the instants of large spurious forces.
Moreover, the spurious forces are on average isotropic.
Benchmark measures that look only at quantities averaged over long times or many particles will therefore miss this error since its mean vanishes.
It can nevertheless contribute significantly to particular configurations where the forces become anisotropic such as near interfaces or in binding pockets, degrading model performance for these locations of high interest.

The symmetrization procedure is straightforward and simply requires rescaling the force field energy parameters.
It can therefore be easily applied and tested in force fields based on Drude oscillators such as the CHARMM Drude force fields \cite{lemkul_empirical_2016}.
The symmetrized force fields are carefully designed to conserve the forces experienced by the atomic centers of mass and should therefore require very little reparametrization of parameters that were fit to only center of mass motions.
However, parameters that were fit to atomic polarization, such as the polarizability tensors, may have had the spurious coupling between polarization and center of mass motion parameterized into them.
For example, in a bonded system, the coupling between the bond potential and atomic polarization along the bond may have been accounted for at a mean field level when the atomic polarizability tensor was parameterized, e.g. by decreasing the polarizability along the bond by an amount that reflects the equilibrium bond length.
Symmetrizing these forces therefore requires additional care and may require a more involved reparameterization but may significantly improve the numerical performance of the force field.

\begin{acknowledgement}
This research was supported under the CPIMS program by the Director, Office of Science, Office of Basic Energy Sciences, Chemical Sciences Division of the U.S. Department of Energy under Contract DEAC02-05CH11231.. A.D. is grateful to Prof. David Limmer for useful discussion.
\end{acknowledgement}

\bibliography{references}

\begin{mcitethebibliography}{52}
\bibitem[Lin and Truhlar(2007)Lin, and Truhlar]{lin_qmmm_2007}
Lin,~H.; Truhlar,~D.~G. {QM}/{MM}: what have we learned, where are we, and
  where do we go from here? \emph{Theoretical Chemistry Accounts}
  \textbf{2007}, \emph{117}, 185--199\relax
\mciteBstWouldAddEndPuncttrue
\mciteSetBstMidEndSepPunct{\mcitedefaultmidpunct}
{\mcitedefaultendpunct}{\mcitedefaultseppunct}\relax
\EndOfBibitem
\end{mcitethebibliography}

\end{document}